\documentclass[12pt,a4paper]{article}
\usepackage{epsf}
\def\COMMENT#1{{\hrule\vskip 0.3cm\noindent\kern -5em\tt*** {#1}
\vskip 0.3cm\hrule}}
\let\epsilon=\varepsilon
\begin{document}
\title { Transport properties and phase diagram of the disordered lattice
vibration model
}

\author { Omri Gat$^{1}$ and Zeev Olami$^{2}$ \\
\normalsize $^1$D\'epartement de Physique  Th\'eorique,
Universit\'e de Gen\`eve\\
\normalsize 32 Bld. D'Yvoy, CH-1211 Gen\`eve, Switzerland.\\
\normalsize $^2$Department of Chemical Physics,
The Weizmann Institute of Science\\
\normalsize
Rehovot 76100, Israel.
}
\maketitle


\begin{abstract}
We study the transport and localization properties of scalar
vibrations 
 on a lattice with random bond strength by means of the transfer
matrix method. This model has been
recently suggested as a means to investigate the vibrations and heat
conduction properties of structural glasses. In three dimensions we find
a very rich phase diagram. The delocalization transition
is split,
so that between the localized and diffusive
phases which have been identified in the Anderson problem, we observe a
phase with anomalous, sub-exponential localization. For low
frequencies, we find a strongly conducting phase with ballistic and
super-diffusive transport, reflecting a diverging diffusivity. The
last phase generates an anomalous heat conductivity which grows with
the system size. These phases are the counterparts of those
identified in an earlier study of the normal modes.

\end{abstract}
  



\bigskip
\section{Introduction}\label{sec:intro}
It is well known that disorder in the parameters of
 physical models can induce scattering and localization, and 
the most outstanding example is the electronic Anderson model. This model
has been the focus
of a large set of theoretical, numerical and experimental studies over
 the last three decades, and there is an enormous literature on
various aspects of
this problem. At the same time, the related problem of 
vibrations in disordered materials  received almost
no attention, although structural disorder is very common
in glasses and other disordered  materials. 

It is well known that in low temperatures there are several features
common to most glasses. An outstanding phenomenon is the existence of an
anomalous heat conductivity at low temperature, a plateau
in the heat conductivity in a higher temperature range, and no decrease in it 
at higher  temperatures~\cite{POHL}. 
It is not clear
if those effects are the results of nonlinear phenomena \cite{PA} or
of harmonic vibrations \cite{GEL86,Allen}. 
A calculation of transport in a vibration
model is a good way of testing whether the main mechanism of heat
conductance in 
glasses is obtained from the linear harmonic vibration modes, or if
nonlinear scattering processes have to be invoked.

Since vibration Hamiltonians are  elastically stable there are
special constraints on the dynamics. Therefore,
though the dynamical equations  
are  
similar to the Anderson problem of a quantum particle in a random
potential there are some major diffrences due to these
constraints. 
This model is a random Laplacian model, with positive off diagonal 
coefficients. It  leads to the existence of a global translation mode
at zero frequency. This features has a decisive influence on the entire
spectrum  of  vibrations at low frequencies. Other distinguishing
features are  the vector
character of the vibrations, and in the case of a two (or more)
component systems, random masses. We will comment in more detail on
these differences in section 2.

The focus of this study is a scalar lattice vibration model with
random elastic coefficients, and equal masses. This model disregards
some of the important features of glass vibrations. Nevertheless,
in a previous work which focused on
the eigenstates \cite{paper1}, it was
shown that this model contains many of the interesting features of
glass vibrations, and is qualitatively different from the usual electronic 
models. The present work reveals the phase diagram and the conductance
properties of the disordered lattice model. We note that the model
under discussion
also describes a free quantum particle with disorder in its transport
behavior.
 
In our study we find several new qualitative  features that are absent
in the classical Anderson problem. Some theoretical studies were
 vibration problems with 
disorder, such as \cite{theory1,theory2}, have
concluded that 
they belong to the same universality class
as the Anderson model. This is 
not the case for in the model studied here and in \cite{paper1}.

\begin{figure}\epsfxsize=15cm
\epsfbox{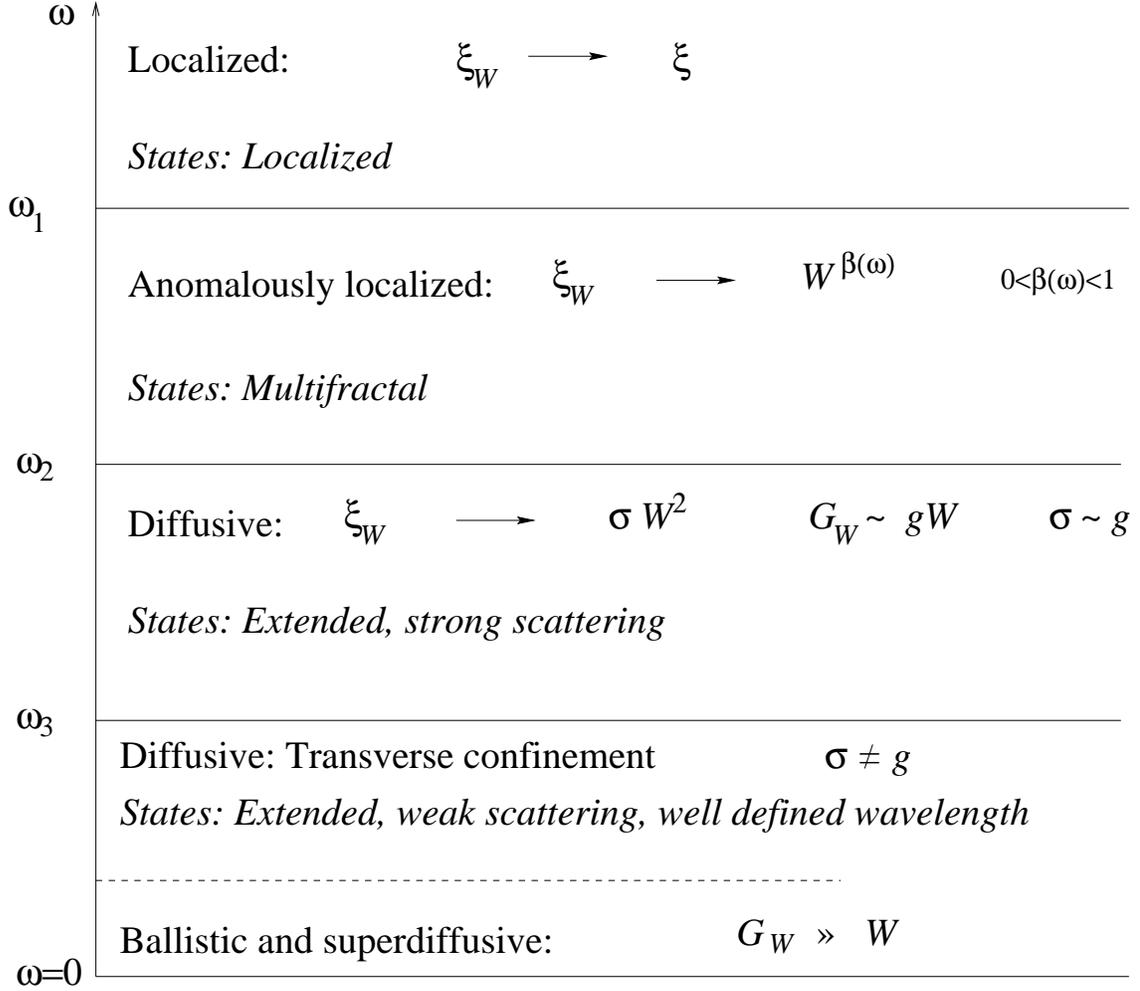}\caption{\label{fig:phase}
Phase diagram of the vibration model in three dimensions: Regions of
different qualitative transport behavior, as a function of the frequency
$\omega$, for a fixed disorder distribution. The small $\omega$ region
below the dashed line is present only in finite systems. The
descriptions given in {\it italics} refer to normal modes, analyzed
in~\protect\cite{paper1}. The symbols
appearing in the diagram are the
finite size localization length $\xi_W$ for a system of width $W$, the
conductance $G_W$ of a finite cubic system of side $W$, and the
conductivity $g$. They are explained in detail in section~\ref{sec:3d}.}
\end{figure}

We outline the main differences:
In the three dimensional electronic Anderson model there are two
phases: diffusive with extended states and localized.
In the vibrational case  there are four phases.
In addition to the two
phases familiar from the Anderson problem there is
an insulating sub-exponentially localized phase with
multifractal normal modes, and a strongly conducting phase, with
states characterized by weak scattering, and a well defined
wavelength. The phase diagram is displayed in fig.~\ref{fig:phase}, and
described in detail in section~\ref{sec:3d}.

In two dimensions the Anderson 
model is insulating with exponential localization and all the states
are multifractal, while in the vibrational case exponential
localization is also observed, but a transition occurs between
multifractal and extended modes, accompanied by a singularity in the
localization length. This is discussed in detail in our paper on the 
two dimensional case \cite{GBO299}.

The former study \cite{paper1} concentrated on a numerical study of
the properties of the normal modes.  In the present paper we apply the
transfer matrix method~\cite{transfer} to the vibration model. This method has
the advantage of pointing out precisely the transition between
conducting and insulating behavior through localization. The finite
size scaling analysis which has been used earlier to calculate the
physical quantities now reveals the presence of additional length
scales. The method also provides a direct tool for calculating the
conductance of finite systems. The various numerical methods which
are used to analyze the Anderson model, and the results obtained have
been recently reviewed by \cite{jr}.

The structure of the paper is as follows:
in section~\ref{sec:model} we define the model, the numerical method,
and explain how the numerical data are used to study the system.
Section~\ref{sec:3d} describes in detail the results in the three
dimensional case, emphasizing the different qualitative
behaviors. In section~\ref{sec:kappa} we apply the results of
section~\ref{sec:3d} to calculate the thermal conductivity as a
function of temperature. The paper is concluded with a
discussion of the relevance of our results to  glasses.

\section{The disordered harmonic model and elastic energy transport}
\label{sec:model}
The vibration problem in glasses is a vector problem which is
defined for a disordered lattice where both structural and elastic
disorder play a major part. In this paper we treat a simpler variant
of this problem: the scalar vibration model. 
The model under consideration is defined on a $d$-dimensional simple cubic
lattice with periodic boundary conditions. It consists of a scalar
field which is subject to elastic equations of motion. For a normal
mode of frequency $\omega$ the field equations read
\begin{equation}\label{de}
-\omega^2\phi(\vec x)=\sum_{y,\,|\vec y-\vec x|=1}k_{\vec x,\vec y}
[\phi(\vec y)-\phi(\vec x)]\ ,\end{equation}
where $\vec x$ and $\vec y$ denote lattice sites.
The $k$'s are symmetric ($k_{\vec x,\vec y}=k_{\vec y,\vec x}$),
 positive, independent and identically distributed
random numbers. A similar model was introduced recently by  
\cite{Sch} to describe boson peaks in glasses. 

This model is in fact a  realization of the Anderson tight
binding model with special constraints.
 The diagonal terms are the sum of the off diagonal terms in the same
row, and the off diagonal terms are real and positive number. The
same model describes free particle.

The focus of the present study is the transport properties of this
model. For this purpose it is convenient to single out a preferred
direction, and label the coordinate in this direction by $t$. We
define a column vector $v_t$ whose components are the values of $\phi$
at the sites on the hyperplane labeled by $t$. It is then possible to
use eq.~(\ref{de}) to express $v_{t+1}$ in terms of $v_t$ and
$v_{t-1}$. This linear relation defines a
random transfer matrix $T$ by
\begin{equation} \pmatrix{v_{t+1}\cr v_t}=T_t\pmatrix{v_{t}\cr v_{t-1}}
\ .\end{equation}
The transfer matrices $T$ have the same non-zero elements as those of
the Anderson tight-binding model. 

Following the standard methods developed in the context of disordered
electronic systems, we study the conductance properties using the
Lyapunov spectrum of the product
\begin{equation}\label{product} \cdots T_tT_{t-1}\cdots T_1\ .\end{equation}
Due to the left-right symmetry of the problem, the Lyapunov exponents 
come in pairs of opposite signs.
For any positive value of $\omega^2$, and for non-vanishing disorder
all the exponents are non-zero, signaling that in a quasi-1D geometry
the system is always insulating. The smallest of the non-zero exponents
may be identified with the best conducting channel, so that it gives
the inverse of the finite size localization length $\xi_W$, where $W$
is the transverse size of the system.

We study $\xi_W$ as a function of $W$, with the aim of finding the
asymptotic behavior for large $W$. The following types of asymptotics
are known to occur in models of electronic localization
\begin{enumerate}
\item The localization length saturates to a finite value 
$\xi_W\rightarrow\xi_\infty=const\ .$
This is the localized phase which is expected to characterize systems
with strong disorder, and low dimensionality.
\item The critical state, $\xi_W\sim W$. This behavior is commonly
associated with the delocalization transition, using the scaling
hypothesis.
\item The diffusive phase, $\xi_W \sim D W^{d-1}$. This phase is
expected to have Ohmic behavior, 
\begin{equation} G\sim g W^{d-2} \ ,\label{gdiff}\end{equation}
and
the conductivity $g$ is proportional to $D$,
\end{enumerate}
Since the transverse size of the system which can be studied
numerically is limited, the extrapolation to infinite size can be
improved significantly using finite-size scaling (FSS) \cite{transfer},
which assumes a scaling form
\begin{equation}\label{fss} \xi_W=\xi_\infty f(W/\xi_\infty)\ ,\end{equation}
with different scaling functions above and below the transition. The
localization length $\xi_\infty$ is found numerically to diverge as
\begin{equation} \xi_\infty\sim |\sigma-\sigma_c|^{-\nu}\ ,\end{equation}
in three dimensions and as 
\begin{equation} \xi_\infty\sim\exp\left(\sigma^{-\nu'}\right)\ ,\end{equation}
in two dimensions where $\sigma$ measures the disorder strength, and
$\sigma_c$ is the critical value of disorder in three dimensions.

The average conductance $G$ is given,
using the the entire set of Lyapunov exponents $\gamma_j$, by
\begin{equation}\label{gl} G=\sum {1\over\cosh^2(\gamma_jW)}\ ,\end{equation}
for a hypercubic sample \cite{review}. We will show below that in the
diffusive state,
since the spacing
between the Lyapunov exponents is slowly varying, the conductivity $g$
is indeed
proportional to $D$. However, this is not always valid, and we show
below that the
relation between $g$ and $D$ breaks down when scattering is too weak.

Here, we study the transport properties for different values of
$\omega$, rather than disorder. As expected, we find that small
$\omega$ corresponds in many ways to weak disorder; the basic reason
for this is that the uncorrelated disorder is averaged out
in its effect on large wavelength modes, which correspond to small
$\omega$. However, the mapping between small $\omega$, and weak noise
is exact only in one dimension~\cite{oned}. We find that the behavior
as a function of $\omega$ is 
considerably more involved than that which is known for electronic
models, and which has been reviewed above. We find several new phases,
and conclude that in three dimensions the simple scaling picture of
the localization
transition does hold for our model. As a result, FSS does not hold in
the simple form~(\ref{fss}), and has to be modified. 
The present study confirms and adds further information to 
the new features which were discovered in the geometry of the eigenmodes
of the same model \cite{paper1}.

\section{Transport properties in three dimensions}\label{sec:3d}
As described is section~\ref{sec:model}, the various phases of the
system can be characterized by the asymptotic behavior for large
transverse width $W$
of the the finite size localization length $\xi_W$ which is the
inverse of the smallest non-negative Lyapunov exponent and of the
conductance $G$ which is defined in terms of the Lyapunov exponents
in~(\ref{gl}). The results of this section were obtained for
the model~(\ref{de}) in a bar shaped geometry, with a
square cross section of width $W$, and a fixed
distribution of the values of the elastic constants $k_{\vec x,\vec y}$: 
A binary distribution with ${\sf Prob}(k=0.1)=0.8$ and ${\sf Prob}(k=1)=0.2$. 
We computed the Lyapunov exponents of the product~(\ref{product}) for
different values of $W$, $5\le W\le16$. The number of realizations
which were multiplied was between $4\times10^6$ for the smallest
widths, and $2.5\times10^5$ for the largest.

We studied the transport properties for varying $\omega$. 
These results are summarized in the
phase diagram displayed
in fig.~\ref{fig:phase}.
We sketch the major features of 
the phase diagram.
For frequencies which are higher then $\omega_1$ we observe a normal
localized phase with exponential localization.
However in the transition range near $\omega_1$, the usual finite size
scaling ansatz fails, and a
modified finite size
scaling with two parameters is needed to rescale the data. 
For smaller frequencies in the range
$\omega_2<\omega< \omega_1$ we observe a divergence of the finite
localization length as a non-trivial power of the system size in the
entire range
available to us. This phase is interpreted as being sub-exponentially
localized. 
In the range $\omega_3< \omega < \omega_2$ we observe the normal diffusive
behavior familiar from electronic models, but  finite size scaling has
again to be modified near the transition.
These effects are related to a continuous 
change in the multifractal 
exponents of the {\it normal modes} in this range, instead of a sharp
transition. 

At frequencies which are 
lower than $\omega_3$, the normal diffusive behavior is
replaced by an irregular dependence of the transport properties on the
system parameters. The diffusion coefficient becomes hard to define
numerically, and a variety of types of size dependence of the conductance is
observed. Since $\omega<\omega_3$
 is the range corresponding to weak scattering as
found in~\cite{paper1}, these effects can be interpreted as residuals of the
ballistic propagation between scattering events. We expect a long
crossover region before an ultimately diffusive behavior can be observed.

%
%

The rest of this section is devoted to a detailed description of
the phases in the order they appear with decreasing $\omega$.

\subsection{The normal localized phase}\label{sec:3dloc}
The system is localized for values of $\omega^2\ge\omega_1^2=1.1$. This
phase is
characterized by the existence of a finite limiting localization
length
\begin{equation} \xi_\infty=\lim_{W\rightarrow\infty}\xi_W<\infty\ ,
\end{equation}
which we conventionally identify with the range of the Green's
function of eq.~(\ref{de})
\begin{equation} \xi_\infty=\lim_{|\vec x-\vec y|\rightarrow\infty}
-{1\over|\vec x-\vec y|}\left<\log|{\cal G}(\vec x,\vec y;\,\omega)|\right>
\ .\label{greens}\end{equation}

We found that in the localized phase the dependence of $1/\xi_W$ on
$1/W$ is well approximated as linear. This function can then be
linearly extrapolated to $1/W=0$, giving an estimate for $1/\xi_\infty$.
The values of $\xi_\infty$ thus obtained are listed
in table~\ref{tab:3dloc} for various values of $\omega^2$.

\begin{table}
\begin{tabular}{c|ccc}
$\omega^2$&		$\xi_\infty$&	$\log_{10}a$&	$\log_{10}b$\\
2.0	&		0.377	&		&		\\
1.6	&		0.131   &		&		\\
1.4	&		0.226	&	-0.16	&		\\
1.3	&		0.259	&	-0.21	&		\\
1.25	&		0.160	&	-0.02	&		\\
1.2	&		0.094	&	\ 0.22	&		\\
1.17	&		0.073	&	\ 0.33	&		\\
1.15	&		0.047	&	\ 0.56  &		\\
1.14	&		0.033	&	\ 0.73	&		\\
1.135    &            	0.026  	&	\ 0.85  &		\\
1.13	&		0.019	&	\ 0.97	&		-0.003\\
1.125	&		0.013   &	\ 1.11	&		-0.007\\
1.12	&		0.008   &	\ 1.20	&		-0.015	\\
1.115	&			&	\ 1.30	&		-0.022\\
1.112	&			&	\ 1.37	&		-0.026\\
1.11	&			&	\ 1.37	&		-0.028\\
1.105	&			&	\ 1.44	&		-0.031
\end{tabular}

\caption{\label{tab:3dloc}
The values of the localization length $\xi_\infty$ as determined by
extrapolation to $W=\infty$, and the parameters $a$ and $b$ of the
modified FSS~(\ref{fss}), as a function of $\omega^2$ in the localized
phase in three dimensions. The values missing from the table could not
be determined significantly numerically, except when a value is given for $a$
and not for $b$, where $b$ was fit to 1. The different symbols were
obtained for different values of $\omega$, all of which are given in
table~\ref{tab:3dloc}. 
}\end{table}

As can be seen from table~\ref{tab:3dloc}, the value $\xi_\infty$
increases as $\omega^2$ decreases as it starts to diverge near the
transition. However, because of the limited size of the systems which
were measured, near the transition the precision of the measured
$\xi_\infty$ deteriorates, until all precision is lost. As discussed
in section~\ref{sec:model} above, this difficulty
is commonly dealt with by invoking the FSS hypothesis~(\ref{fss}),
which can be written equivalently
as $\log(\xi_W/W)=g(\log W-\log\xi_\infty)$; $\xi_\infty(\omega)$ is
then determined so that the various curves of measured $\log(\xi_W/W)$
versus $\log W$ collapse.

\begin{figure}\epsfxsize=15cm
\epsfbox{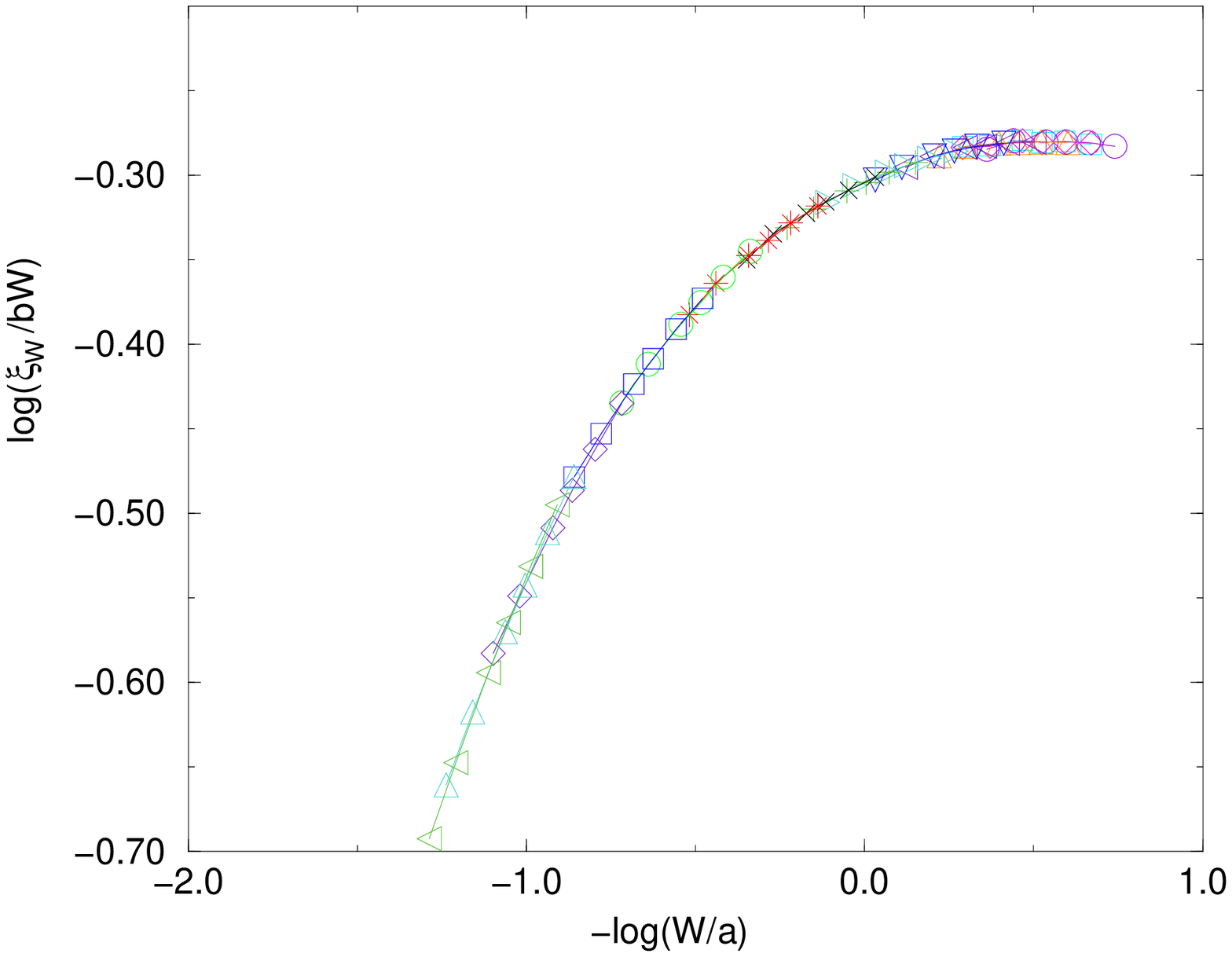}\caption{\label{fig:3dloc}Modified
finite size scaling in the localized phase in three dimensions. $W$ is
the width, and $\xi_W$ is the finite size localization length; the
logarithms are in base 10. The
data displayed are for
$1.105\le\omega^2\le1.4$. The FSS parameters
$a$ and $b$ as functions of $\omega$ are
reported in table~\ref{tab:3dloc}}
\end{figure}

This procedure works well for $\omega^2\ge1.14$. However, for
$\omega_1^2=1.1\le\omega^2\le1.14$ the numerical curves do not
collapse implying
that FSS is violated.
It is nevertheless possible to describe the measured values of $\xi_W$
for all $\omega^2$ using a
modified form of FSS
\begin{equation} {\xi_W\over W}=b(\omega^2)f\left({W\over a(\omega^2)}\right)
\label{modfss}\ ;\end{equation}
the values of $\log a$ and $\log b$ thus obtained are also reported in
table~\ref{tab:3dloc}, and the resulting data collapse is displayed in
fig.~\ref{fig:3dloc}. Unfortunately the fitting parameters $a$ and
$b$ are not directly related to the localization length, and therefore
the modified FSS is not useful for a measurement of $\xi_\infty$ near
the transition. The physical interpretation of $a$ and $b$
 remains unclear.
For $\omega^2\ge1.14$, where $b=1$, $a$ may be again identified with
$\xi_\infty$, which gives values in accord with those obtained from
direct extrapolation.

As discussed above, $\xi_\infty$ diverges at the transition, but it is
difficult to measure this divergence accurately. Nevertheless, the
values of $\xi_\infty$ obtained by direct extrapolation are
well in accord with a power like divergence
\begin{equation}\label{nu1}\xi\sim(\omega-\omega_1)^{-\nu_1}\end{equation} 
with $\nu_1\sim1.7$. This value
is within the numerical errors in agreement with the value of
$\nu=1.5$ reported for the three-dimensional Anderson
transition~\cite{review}. 
On the other hand, the fitting parameters $a$ and~$b$ increase very
slowly near $\omega_1$, and their rate of divergence was impossible to
determine to a reasonable accuracy, from the numerical measurement.
The breakdown of FSS implies
 there is at least one more
length scale, $\xi_*$, in the system in addition to $\xi_\infty$(and $W$);
unlike $\xi_\infty$, $\xi_*$ does not diverge at the
transition.

\subsection{The anomalous localized phase}
\begin{figure}\epsfxsize=15cm
\epsfbox{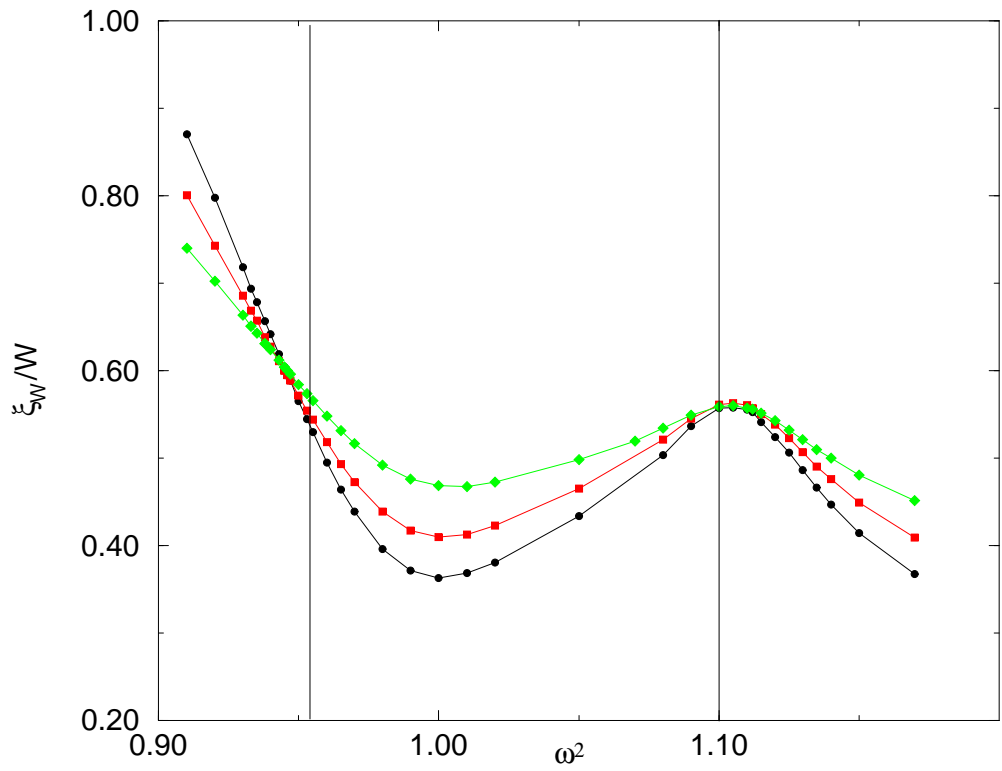}\caption{\label{fig:scanaloc}Finite
size localization length $\xi_W$ divided by the width $W$, as a function of
the frequency squared $\omega^2$, for widths $W=5,\,8,\,12$, from top
to bottom.
The vertical lines show the limits
of the anomalous
localized phase.}
\end{figure}

This subsection describes the properties of the system for
$\omega_2^2=0.96\le\omega^2\le\omega_1^2=1.1$. In electronic models
the Anderson 
transition in three dimensions is the boundary between localized
and conducting diffusive phases. In the present model the
qualitative behavior is markedly different. One surprising feature
is that the finite size localization length $\xi_W$ for a given $W$
viewed as a function of $\omega^2$ has a {\em maximum} at the the transition
point $\omega^2=1.1$. That is, the conductivity of a finite sample
becomes worse when $\omega$ is decreased from the transition,
the opposite of what would occur if the transition were
into a conducting phase. This trend continues until a minimum is
reached at $\omega^2=1.0$, and the usual situation whereby $\xi_W$
increases with decreasing $\omega^2$ is restored (see
fig.~\ref{fig:scanaloc}).

\begin{figure}\epsfxsize=15cm \epsfbox{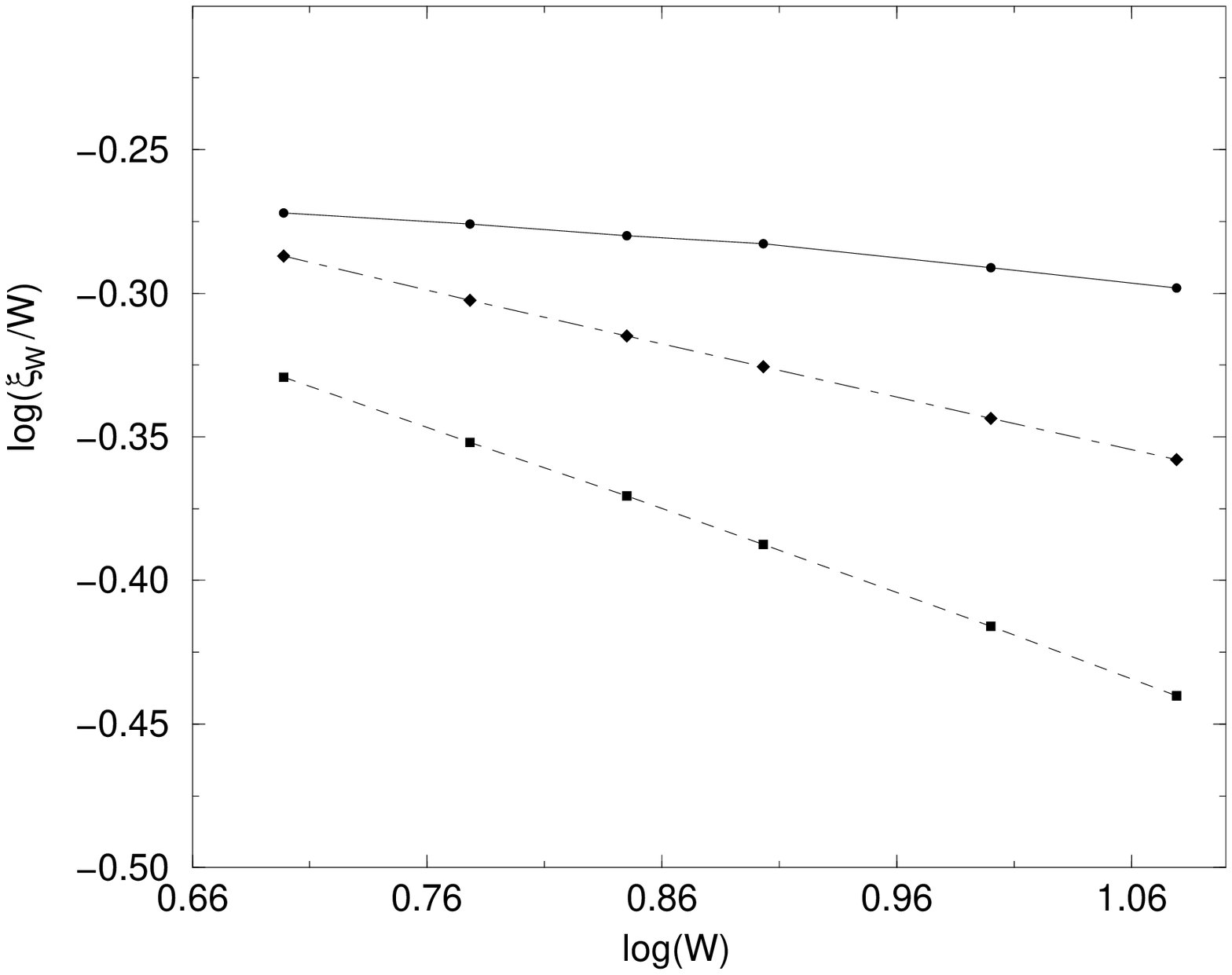} \caption{
\label{fig:aloc}The base 10 logarithm
of the Finite
size localization length $\xi_W$ divided by the width $W$, versus
$\log_{10}W$, shown for three frequencies in the anomalous localized
range.
The full, dashed, and dot-dashed
lines refer to $\omega^2=1.08,\,1.00,\,0.97$ respectively.}\end{figure}

The behavior for a fixed $\omega$ as a function of $W$ is also
unusual. $\xi_W$ does not saturate to a finite value, and continues to
grow with $W$ as some non-trivial power
\begin{equation} \label{xibeta}
\xi_W\sim W^\beta\ ,\qquad0<\beta<1\ .\end{equation}
Some examples are given in fig.~\ref{fig:aloc}, where a clear power
law behavior is observed, in contrast with the other phases.

The measured values of $\beta$ are reported in table~\ref{tab:aloc},
and the minimum occurs at $\omega^2=1$. An interesting feature is that
the at the lower bound of this phase $\beta$ is still significantly
smaller than 1; we discuss below the implication of this fact on the
nature of the second phase transition into the diffusive phase.
\begin{table}
\begin{tabular}{c|cc}
$\omega^2$&$\beta$&$\xi_*$\\
1.10&	1.00&\\
1.09&	0.97&	0.93\\
1.08&	0.93&	0.95\\
1.05&	0.84&	0.87\\
1.02&	0.74&	0.98\\
1.01&	0.73&	0.95\\
1.00&	0.71&	1.00\\
0.99&   0.72&	1.03\\
0.98&	0.75&	1.08\\
0.97&	0.81&	1.15\\
0.965&	0.85&	1.25\\
0.96&	0.88&	1.40\end{tabular}
\caption{\label{tab:aloc}The localization exponent $\beta$ and the
anomalous localization length $\xi_*$ as a function of $\omega^2$, in
the anomalous localized phase.}
\end{table}
The asymptotics of $\xi_W$ given in eq.~(\ref{xibeta}) shows that the
Green's function in this phase is certainly not localized with a
finite localization length $\xi_\infty$ in the sense defined
in~(\ref{greens}). However, it is compatible
with a weaker `stretched exponential' localization
\begin{equation} \label{strloc}\left<|{\cal G}({\vec x},{\vec y})|\right>\sim
\exp\left[-\left({|{\vec x}-{\vec y}|\over\xi_*}\right)^\alpha\right]
\ ,\end{equation}
with $0<\alpha<1$. If the three dimensional system indeed behaves
as in~(\ref{strloc}) then we can expect the same behavior in the bar
geometry which we studied for distances $|{\vec x}-{\vec y}|$ which
are small compared to $\xi_W$, which marks the crossover to true
exponential decay. If $W\gg\xi_*$, we can estimate $\xi_W$ by
demanding that the two types of decay are of the same order of
magnitude for distances of order $W$, that is,
\begin{equation} \exp\left[-\left({W\over\xi_*}\right)^\alpha\right]\sim
\exp \left[-{W\over\xi_W}\right]\ .\label{cross}\end{equation}
This yields the estimate
\begin{equation} \xi_W\sim W^{1-\alpha}\xi_*^\alpha\ ,\end{equation}
which agrees with~(\ref{strloc}) identifying $\beta=1-\alpha$.


The delocalization at
$\omega\to\omega_1$ from below, as described in this section, occurs
because $\beta\to1$, and the length $\xi_*$ remains finite. It seems
natural therefore to identify $\xi_*$ with the additional length scale
present in the normal localized phase.

The transition in $\omega_2$ is unique since 
$\alpha(\omega_2)>0$. One has a stretched exponential  correlation 
with an internal length $\xi*$ which should diverge to infinity at the
transition.  This is analog to the
 usual process of delocalization. The value of $\xi_*$ can
be calculated from 
the numerical results using
eq.~(\ref{cross}), and the values obtained are reported in
table~\ref{tab:aloc}. However,  the measured
values of $\xi_*$ are  inaccurate, and should be considered
only as indicative. If these values are trusted, the conclusion
is that the divergence of $\xi_*$ near $\omega_2$ is very slow, but a
quantitative statement is impossible to make.

In the analysis of~\cite{paper1} the normal modes in parameter range
described in this 
section could not be observed with a good enough resolution. 
A more detailed study enabled us
 to observe this range with better
statistics, and the conclusion is that the states are multifractal in
this range, with
exponents changing continuously, in a way similar to the change of
the exponent $\beta$ of the divergence of the
correlation length. These results and the relationship between them
and the present study will be given in a separate paper \cite{GBO299}.
 A possible way to describe the phase described in this section, is
that it is an intermediate regime, where the change in the fractal 
exponents smears the Anderson transition.
%

\subsection{The diffusive phase}

This is a conducting phase with Ohmic behavior, observed in the
parameter range 
$\omega_3^2=0.45\le\omega^2\le\omega_2^2=0.96$. 
The characterizing feature of this phase is the existence of a
well-defined nonzero limit
\begin{equation} \label{diff} \lim_{W\rightarrow\infty}
{\xi_W\over W^2}=D\ .\end{equation}
$D$ is approximately proportional to the conductivity
of the system, as is discussed in detail below.
The lower limit of this phase $\omega_3$ is not very well defined,
because finite size effects are stronger for small $\omega$.

As in the normal localized phase (see subsection~\ref{sec:3dloc})
there are two methods in principle to calculate $D$ from the raw
numerical data. One may regard $D_W=\xi_W / W^2$ as a function of
$1/W$ and extrapolate to $1/W=0$, which should work well far from the
transition point $\omega_2$, where $D$ is not too small, or, near the
transition $D$ may be inferred from FSS, eq.~(\ref{fss}), where in this case
the scaling function $f$ grows quadratically for large
arguments~\cite{review}. However, as in the normal localized phase,
the standard form of FSS breaks down near the transition and has to be
replaced by the modified FSS~(\ref{modfss}), which is not useful for
the determination of $D$. Again, this breakdown implies the presence
of additional length scales.

Since FSS is of limited use, the measured values of $D$
reported in table~\ref{tab:3ddiff} were obtained by straightforward
extrapolation. The values approach zero when
$\omega\rightarrow\omega_2$ as in the usual Anderson transition. The
decay rate can be fit quite convincingly with power law 
\begin{equation}D(\omega)\sim(\omega_2-\omega)^{\nu_3}\ ,\end{equation}
with $\nu_3 = 1.5 \pm 0.1$,
 again in agreement with the Anderson transition, 
within numerical error with $\nu_1$ which measures the divergence of
the localization length near the boundary between the normal localized
phase, and the anomalous localized phase [see eq.~(\ref{nu1})].
\begin{table}
\begin{tabular}{c|cccc}
$\omega^2$&		$D$&	$\log_{10}a$&	$\log_{10}b$&$g$\\
0.50&			0.891&		0.786\\
0.60&			0.657&	0.00&&	0.554\\
0.70&			0.423&	0.20&&	0.354\\
0.75&			0.283&	0.35&&	0.245	\\			
0.80&			0.173&	0.57&&	0.148	\\	
0.82&			0.127&	0.68&&	0.115\\		
0.85&			0.076&	0.90&&	0.070\\		
0.87&			0.050&	1.08&&	0.047\\				
0.89&			0.032&	1.29&&	0.030\\
0.90&			0.025&	1.40&&	0.024\\		
0.91&			0.019&	1.55&&	0.018\\
0.92&			0.014&	1.74&&	0.012\\
0.93&			0.008&	2.03&&	0.005\\	
0.933&			0.006&	2.10&			0.005&	0.004\\
0.935&			&	2.18&			0.008&\\
0.938&			&	2.28&			0.014&\\
0.94&			&	2.36&			0.019&\\
0.943&			&	2.49&			0.028&\\
0.945&			&	2.60&			0.036&\\
0.946&			&	2.68&			0.041&\\
0.947&			&	2.71&			0.045&\\
0.95&			&	2.90&			0.063&\\
0.953&			&	3.14&			0.088&\\
0.955&			&	3.29&			0.107&
\end{tabular}
\caption{\label{tab:3ddiff}
Same as table~\ref{tab:3dloc} for the three-dimensional diffusive
phase, except that  $D$ is given instead of
$\xi_\infty$, and the conductivity $g$ is also given. It is evident
that $D$ and $g$ are close throughout this phase.
}\end{table}

\begin{figure}\epsfxsize=15cm
\epsfbox{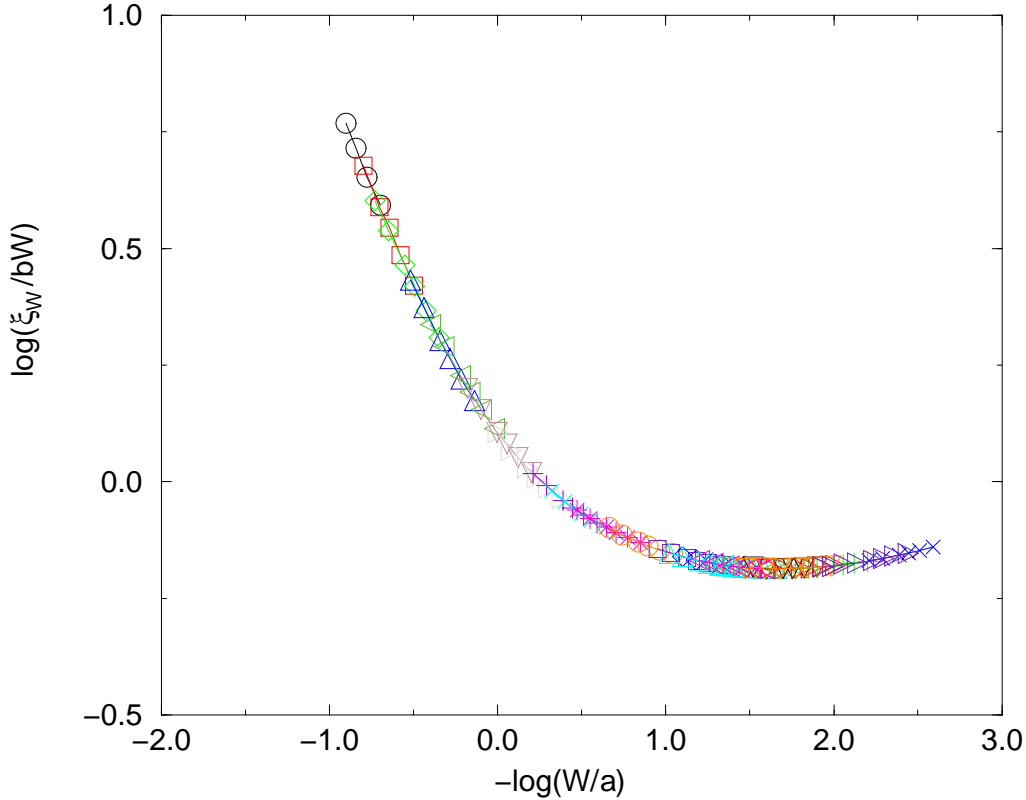}\caption{\label{fig:3ddiff} Same as
fig.~\ref{fig:3dloc} for the diffusive phase in three dimensions. The
frequency range is $0.5\le\omega^2\le0.955$.
$a$ and $b$ as functions of $\omega$ are
reported in table~\ref{tab:3ddiff}}
\end{figure}

The values of $\log a$ and
$\log b$ obtained from the modified FSS ~(\ref{modfss}) are also reported in
table~\ref{tab:3ddiff}, and the data collapse obtained after scaling
the raw data is shown in
fig.~\ref{fig:3ddiff}. For values of $\omega$ where $b=1$, the
parameter $a$ may be identified with $\xi_\infty\equiv 1/D$, with good
agreement with the values of $D$ measured by direct extrapolation.

As can be seen from table~\ref{tab:3ddiff},
the parameters $a$ and $b$ increase in absolute value as
$\omega$ is increased toward $\omega_2$, and
diverge slowly. This implies an interesting
behavior of the localization properties near $\omega_2$: Suppose that
the asymptotic behavior of the scaling function $f$ in~(\ref{modfss})
for small argument is power like,
\begin{equation} f(x)\sim x^\beta\ ,\qquad\mbox{for $x\rightarrow0$}\ .
\label{scfnas}\end{equation}
Taking a value of $\omega$ close to $\omega_2$, so that $a$ is very
large, one finds from eqs.~(\ref{modfss}) and~(\ref{scfnas}) that 
\begin{equation} \xi_W\sim W^\beta \label{ntas}\end{equation}
for $W\ll a$. This behavior has to match the one observed at the lower
edge of the anomalous localized phase, which implies the identity
\begin{equation} \beta=\beta(\omega_2)\sim0.88\ .\end{equation}
The dependence on $W$ shown in~(\ref{ntas}) means that very close to
the transition the system looks localized for small $W$ because
$\xi_W/W$ is decreasing, before the crossover to the asymptotic
 linear growth of $\xi_W/W$. This scenario is corroborated
by the minimum in the curve shown in fig.~\ref{fig:3ddiff}.

Since this is a conducting phase, it is instructive to compare the
parameter $D$ with the conductivity $g$. The calculation of the latter
from the Lyapunov spectrum may be achieved using eqs.~(\ref{gl})
and~(\ref{gdiff}). We 
remark that the dimensionless conductance $G$ calculated
from~(\ref{gl}) is sometimes called the transmission coefficient,
since it is obtained by regarding the system as placed between ideal
leads and providing $W^{d-1}$
channels which conduct energy between the leads. The precise
dependence of $G$ on the channel transmissions is model dependent,
and~(\ref{gl}) reflects one possible choice~\cite{channels}.

\begin{figure}\epsfxsize=15cm\epsfbox{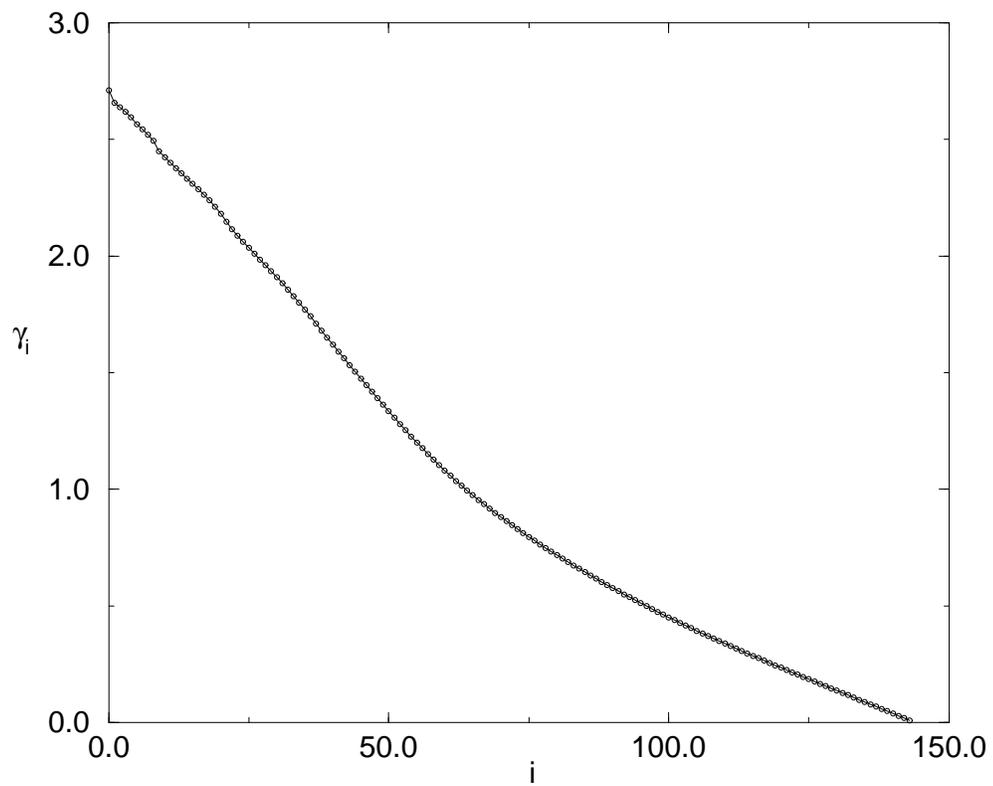}\caption{\label{fig:lyapdiff}
A typical Lyapunov spectrum in the diffusive phase, with the exponents
decreasing linearly to zero. $\gamma_i$, $i=1,\ldots,144$ are the
positive Lyapunov
exponents for a bar of width 12, at $\omega^2=0.6$.}
\end{figure}

The reason that $D$ gives a good estimate for $g$ in the diffusive
phase is that the Lyapunov spectrum in this phase vanishes linearly at
zero \cite{corr}, and the spacing between the Lyapunov exponents
changes slowly,
as exemplified in fig.~\ref{fig:lyapdiff}. This is in contrast with
the Lyapunov spectrum in the low frequency range discussed below. 

To see how $g$ and $D$ are related let us assume that the $W^2$
Lyapunov exponents are evenly spaced, with spacing $1/\xi_W$. The
formula for the conductivity~(\ref{gl}) involves a sharp cutoff at
$\gamma_n\sim W$,  so it basically
counts the number of exponents smaller than $1/W$. Therefore we get 
$G \sim  \xi_W/W$. Since $\xi_W\sim DW^2$ and
 $G=gW$,  $g$ and
$D$ are proportional. In table~\ref{tab:3ddiff} we give the measured
values of $g$. The agreement between $D$ and $g$
remains good throughout the diffusive phase.

\subsection{The strongly conducting phase}
In this subsection we describe the properties of the system for
$\omega^2\le\omega_3^2=0.45$.

We find that for very small $\omega$ short range disorder
becomes irrelevant, and the vibration modes behave very similarly to
modes of a pure system with a well-defined wave vector $k$ and an
acoustic dispersion relation $\omega=c|k|$. Such modes exist
even in one-dimensional systems, but are restricted to a band of width
$W^{-1/2}$ around $\omega=0$.
Similarly in three dimensions, although for any positive $\omega$ the
transport is ultimately diffusive, there is a band of ballistic modes with a
well defined wave vector near
$\omega=0$, which disappears in the thermodynamic limit. The existence
of this band may be simply understood as a consequence of the
divergence of the mean free path $\ell$ as $\omega\to0$. For any finite
system there is a positive frequency $\omega_b$ where $\ell$ becomes
larger than 
the system size, and for smaller frequency a wave propagates through
the system with essentially no scattering.

Outside the ballistic band, and for $\omega\le\omega_3$,  the
normal mode analysis of~\cite{paper1} shows that the  modes
experiences scattering which destroys the unique 
 wave vector dependence. However, the absolute scale of the wave 
vector, the {\em wavelength}, is still well defined, and it is
significantly larger than the lattice
spacing. The wavelength is defined up to some width,
 which is commonly
associated with the inverse of the mean free path.

\begin{figure}\epsfxsize=15cm\epsfbox{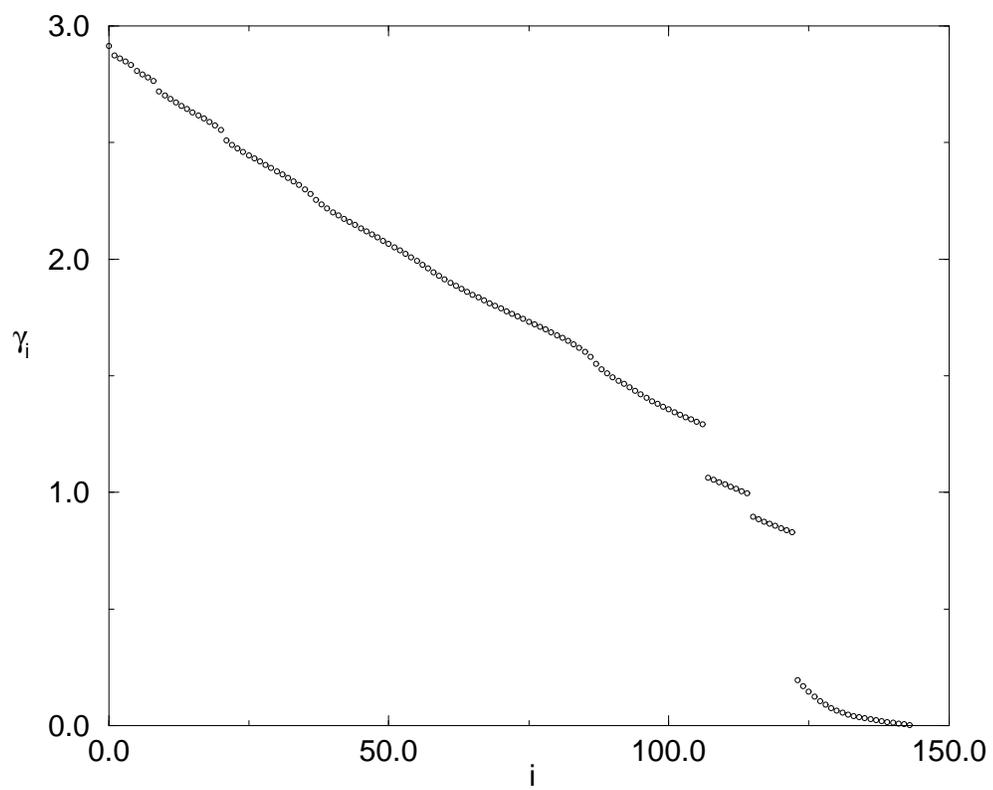}\caption{\label{fig:lyapbal}
A typical Lyapunov spectrum in the strongly conducting,
for a bar of width 12, at $\omega^2=0.2$ (compare with
fig.~\ref{fig:lyapdiff}.)}
\end{figure}

The transport analysis of the present study  reveals that a qualitative
change in the structure of the
Lyapunov spectrum  occurs at $\omega_3$. 
In fig.~\ref{fig:lyapbal} a typical Lyapunov
spectrum of the low frequency phase is displayed, to be compared
with fig.~\ref{fig:lyapdiff}. The Lyapunov spectrum at small $\omega$
is characterized by the presence of large gaps; these obviously rule
out the possibility of a
description of the conductivity using a single parameter. The
conductivity of the system  depends sensitively on the system size,
as well as the other parameters of the problem.

The origin of the gaps in the Lyapunov spectrum can be traced to
the behavior of a pure system which is extended infinitely
in the longitudinal direction. In such a system, for
every longitudinal wave number $k_\parallel$, there are $W^2$ modes
with 
\begin{equation}\label{cs}
\omega_n^2=c_s^2(k_\parallel^2+k_n^2)
\end{equation}
where $k_n$, $1\le n\le W^2$
are the possible values of the transverse wave vector. These
represent $W^2$ channels, which 
are perfectly conducting in the pure system.  However, for a given $\omega$
one will have less then $W^2$ solutions since
 some of channels may
be blocked, when the transverse vector is too large (transverse confinement).
 Therefore the
conductivity of the pure system depends discontinuously on $\omega$:
When $\omega$ is increased beyond the threshold values the
conductivity jumps by an integer amount.


This picture describes quite accurately the conductance $G$ for
small frequencies $\omega\ll\omega_b$. As $\omega$ is increased the
Lyapunov spectrum is less well described by the one of an effective
pure system, but gaps and irregularities in the spectrum persist up to
$\omega_3$, reflecting a non-vanishing transverse confinement.

Consider the conductance $G$ as a function of $\omega$ for a fixed
system size $W$. As just explained, for small $\omega$
system size the conductance is ballistic, and the value is
proportional to the number of open channels, which is proportional to
$(\omega W)^2$. This behavior persists as long as $W$ is smaller than
the mean free path $\ell$. Since the mean free
path diverges as $\ell\sim\omega^{-4}$ in three dimensions
\cite{theory1,Sch}, the crossover occurs at $\omega_b\sim W^{-1/4}$,
therefore $G(\omega_b)\sim W^{3/2}$. 

Since the system is expected to be diffusive for any positive
$\omega$, the dependence of $G$ on $W$ should change from the
quadratic dependence of the ballistic range to an ultimately linear
one, $G\sim gW$. 
If we assume a simple crossover from ballistic to
diffusive behavior of $G$ when $W\sim\ell$, continuity of $G$ in $W$
implies that $(\omega_b W)^2\sim W^{3/2}\sim g(\omega_b)W$, and
the conductivity would diverge as $g(\omega)\sim\omega^{-2}$. However,
the conductivity should diverge at the same rate as the mean free
path, {\it i.e.}, as $\omega^{-4}$. It follows then that there is a
range in $W$ where $G$ grows stronger than linearly but weaker than
quadratically, before the final diffusive behavior sets in.

\begin{figure}\epsfxsize=15cm\epsfbox{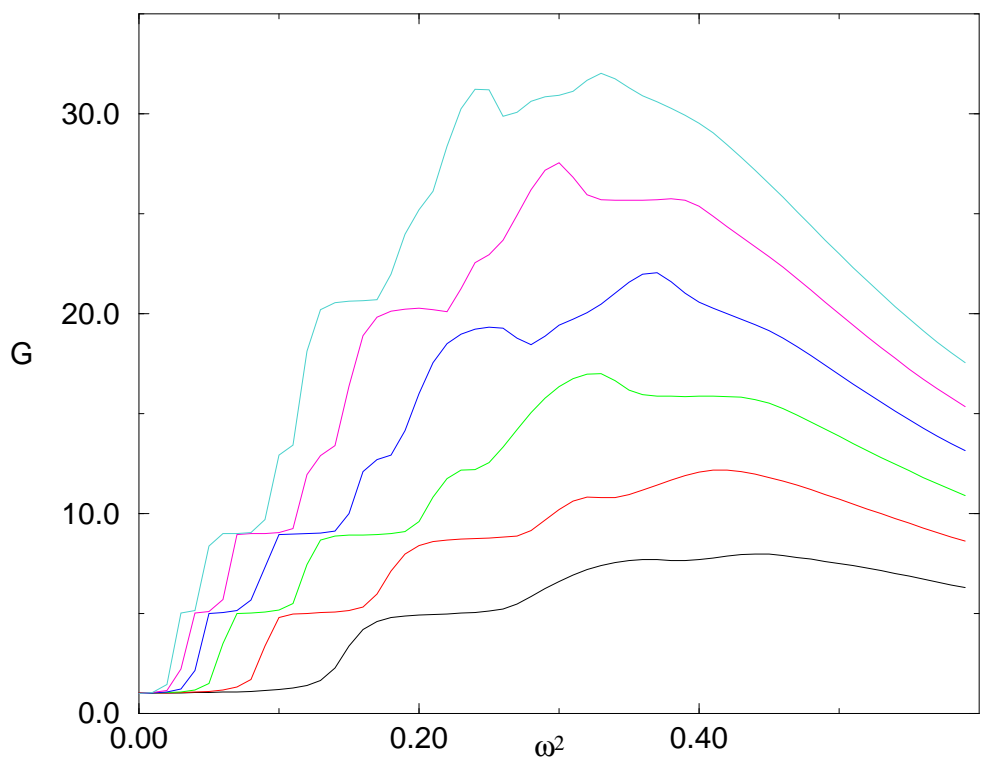}\caption{\label{fig:condcan}
The conductance $G$ as function of the frequency squared $\omega^2$
for different bar 
widths. From bottom to top the widths are $W=6,\,8,\,10,\,12,\,14,\,16$.
The dependence on $W$ for $\omega^2>0.45$ is linear.}
\end{figure}

\begin{figure}\epsfxsize=15cm
\epsfbox{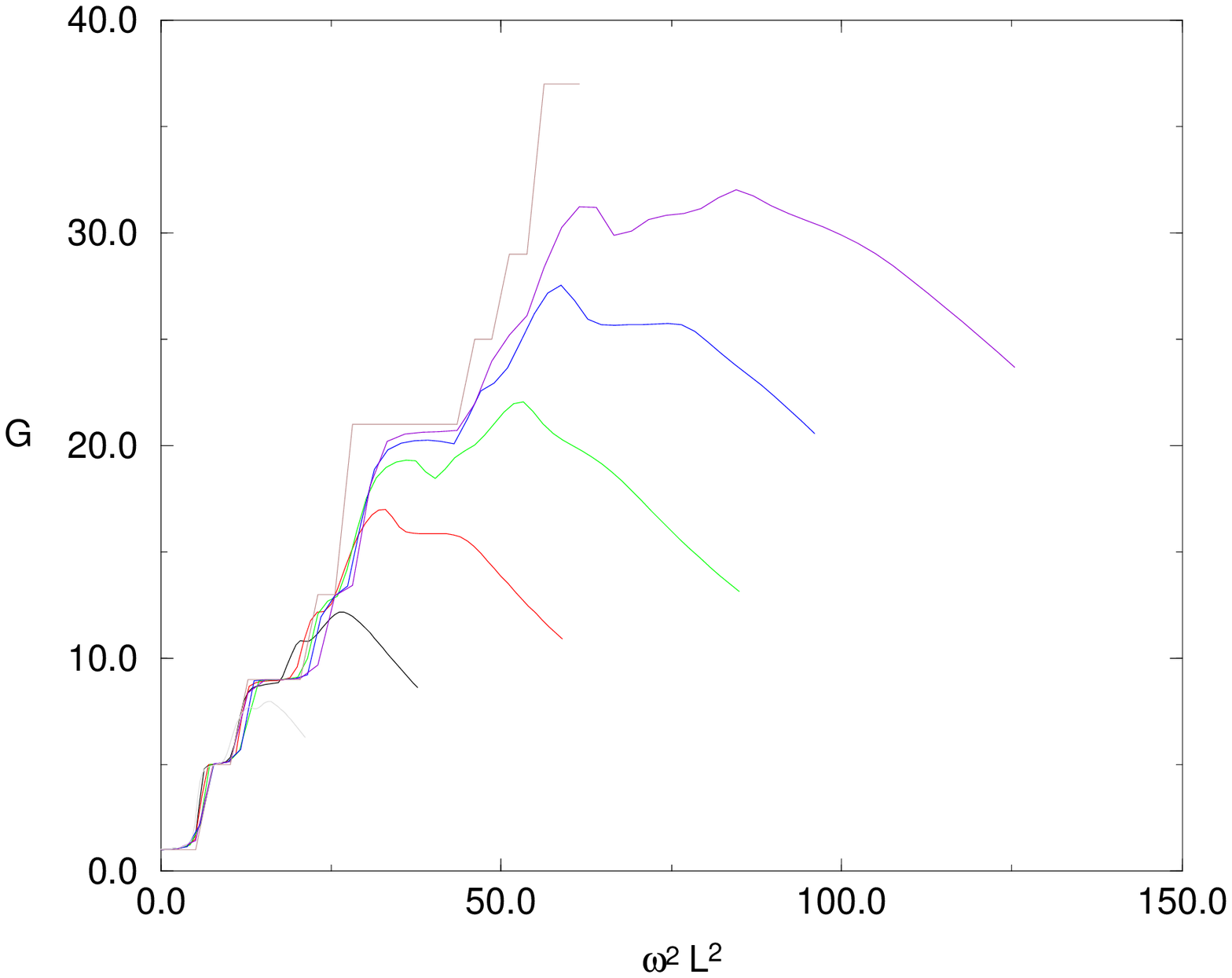}\caption{\label{fig:condcollapse}
The same data as in fig.~\ref{fig:condcan} but shown as a function as
$\omega^2W^2$. The top curve is the result that would be obtained for
a pure system with sound speed $c_s=0.35$.}
\end{figure}

These different behaviors and crossovers are represented in the
conductance as measured numerically, and presented in
fig.~\ref{fig:condcan}. The conductance at small $\omega$ is very well
described by that of an effective pure system with sound speed
$c_s=0.35$, which is between the two possible values for the elastic
constant, $0.1$ and $1$ (recall that the masses were set to 1). The
agreement is quantitative as is demonstrated in
fig.~\ref{fig:condcollapse}, where the measured conductivity is shown against
$\omega^2W^2$, and compared with that of the effective pure
system. At the frequency range $\omega_b\le\omega\le\omega_3$ the
functional form of $G$ is harder to describe. $G_W$ indeed grows
faster than linearly and slower than quadratically, but as a result of
the transverse  effects discussed above, the growth rate is not
monotonic and subject to sharp changes. Thus, we were unable to
measure precisely the growth rate of $G$ as a function of $W$ for a
fixed $\omega$. On the other hand it has been possible to measure the
growth rate of the maximal value of $G$ for a fixed $W$
\begin{equation} \max_\omega G_W(\omega)\sim W^{1.5}\ ,\end{equation}
which is the smallest possible growth rate,
since $\max_\omega G_W(\omega)\ge G(\omega_b)\sim W^{3/2}$. The
ultimately linear growth of $G$ as a function of $W$ could not be observed
in our numerics in this range for our system
sizes.

\section{Thermal conductivity in three dimensions}
\label{sec:kappa}
The results of the previous section can be used to calculate the
overall heat conductivity of a three-dimensional disordered harmonic
lattice, in the presence of a temperature gradient. The heat flow
through a sample connected to two infinite ideal leads at different
temperatures has been
recently derived~\cite{heat}. From this, the thermal conductivity
$\kappa$ may
be calculated, taking the temperature difference to be small, giving
\begin{equation} \kappa={\hbar\over2\pi}{1\over W}\int_0^\infty d\omega
\omega\partial_T\eta_T(\omega)G(\omega)\ ,\label{kappa}\end{equation}
where $G(\omega)$ is the frequency dependent transmission rate, discussed in
the previous section, and
$\eta_T(\omega)=1/(e^{\hbar\omega/T}-1)$. Below we use units such that
Planck's constant and Boltzmann's constant are both equal to one.

The distribution function $\eta_T(\omega)$ which appears in
~(\ref{kappa}) diverges like $T/\omega$ for 
small $\omega$, and has an exponential cutoff at
$\omega\sim T$. Hence for the purpose of a qualitative estimate we can replace
$\eta_T(\omega)$ by $T/\omega$, cutting off the frequency integration
at $\omega= CT$, with $C$ an order 1 constant. This approximation gives
\begin{equation}\label{kps}\kappa\sim {1\over W}\int_0^{CT} d\omega G(\omega)
\ .\end{equation}
Thus, as the temperature increases, modes with higher $\omega$
contribute to $\kappa$; if $CT$ is in the frequency range of one of the
phases, then some of the modes of this phase contribute to $\kappa$,
the phases of lower $\omega$  contribute fully, and those of higher
$\omega$ do not contribute at all.

It follows that when $T\ll\omega_3$ only the strongly conducting phase
should be 
taken into account. If the system  is small enough so that $T\ll\omega_b$,
the conductance is carried by ballistic modes, and eq.~(\ref{kps}) gives
\begin{equation} \kappa\sim WT^3\ ,\end{equation}
as in a perfect lattice. When $T>\omega_b$ this form is no longer
valid, and scattering should be taken into account. Unfortunately, the
numerical results provide limited information about the asymptotic
behavior of $G$ in this range, and we are forced to make an ansatz
about the form of $G$ which takes into account the numerical results
and theoretical predictions, such as
\begin{equation} G\sim
\left\{\begin{array}{cl}\omega^2W^2&0<\omega<\omega_b\sim W^{-1/4}\\
W^{3/2}&\omega_b<\omega<\omega_c\sim
W^{-1/8}\\\omega^{-4}W&\omega_c<\omega<\omega_3
\end{array}\right.\ .\end{equation}
This ansatz gives a linear dependence of $\kappa$ on $T$ in the
intermediate frequency range, with a coefficient proportional to
$W^{3/8}$ and a fast saturation at higher temperatures.

At temperatures greater or of the order of $\omega_3$ the contribution to
$\kappa$ of the strongly conducting phase is therefore {\em anomalous}
growing with the system size as $W^{3/8}$. The exponent of the anomaly
should be considered as only indicative, but the contribution of the
ballistic band itself already gives a $W^{1/4}$ anomaly. The
divergence of the heat conductivity of the harmonic modes at low
frequency is quite robust, and has been observed also in molecular
simulations of glasses \cite{Allen}. Nonlinear interaction are 
invoked  to obtain heat conductivity which remains finite in
the thermodynamic limit. This can be done phenomenologically by
replacing the system size dependence of $\kappa$ by the inelastic mean
free path.

In contrast, the contribution of the diffusive phase is quite
simple:
\begin{equation} \kappa_{\rm diffusive}\sim \int_{\omega_3}^{CT} D(\omega)\ ,
\end{equation}
which gives Ohmic, system size independent, conductivity, increasing
linearly at first, and 
saturating as $T\sim\omega_2$. This contribution, however, becomes
negligible when the system size is taken to infinity. Finally the
contribution of the localized phases to $\kappa$ decays exponentially
in the system size, and these phases are insulating.

\begin{figure}\epsfxsize=15cm\epsfbox{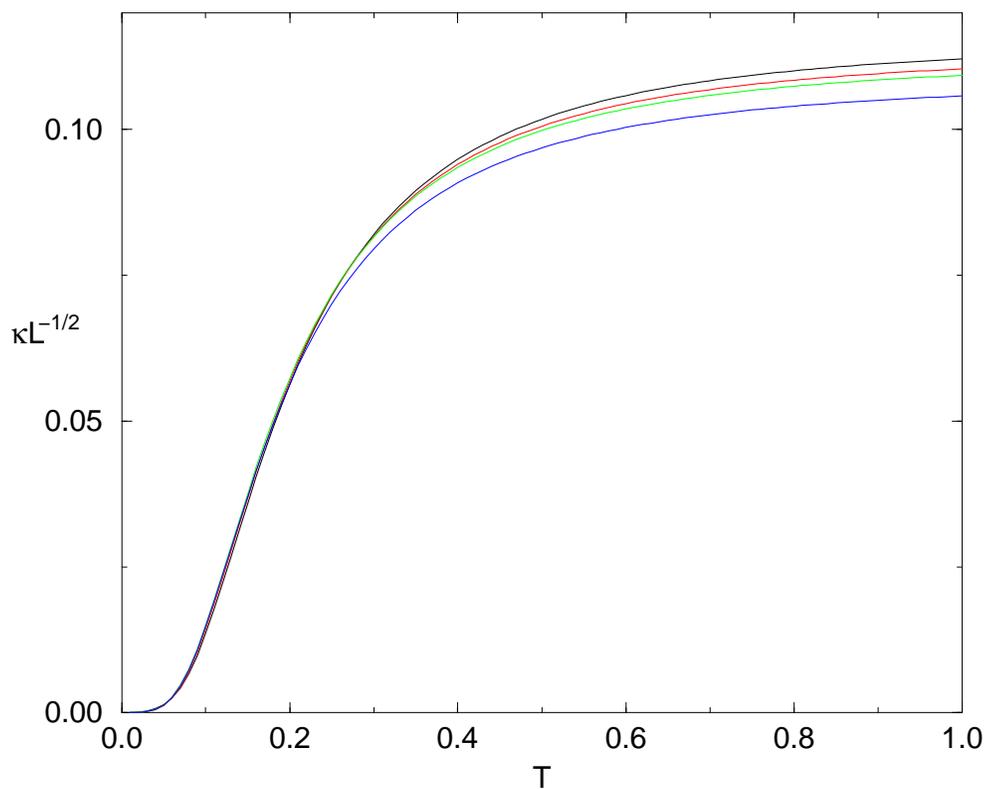}\caption{\label{fig:kappa}
The thermal conductivity $\kappa$ of a cube of side $L$ divided by the
square root of $L$, as a function of the temperature $T$, shown for
different
$L$s. The values of $L$
shown are (from to to bottom) 6, 8, 10, and 12.}
\end{figure}

We also used eq.~(\ref{kappa}) to calculate $\kappa(T,W)$ directly
from the numerical values of $G(\omega)$; the results are displayed in
fig.~\ref{fig:kappa}, scaled with $W^{1/2}$. The integration over
$\omega$ suppresses the finite size effects, and it can be seen that
the $W^{1/2}$ scaling slightly overestimates the measured anomaly. As
expected, one observes in fig.~\ref{fig:kappa} a $T^3$ increase for
small $T$, followed by a linear increase for intermediate $T$ and a
saturation. A power law fit for the saturation yields
an $W^{0.4}$ anomaly, but we regard this value as indicative only.

\section{Conclusions}
We wish to comment on two central issues related to our model. 
First, in the scalar vibration  model the delocalization transition
 is smoothed  in the sense that there are two  critical points,
 and there
is a continuous change of the multifractal exponents near the
transition.
Moreover,  far from the transitions one can observe
the classical finite size scaling behavior similar to 
 the Anderson scaling.

Our second principal conclusion is
that the heat transport of modes of low frequency has some
peculiarities probably related to the existence of a well defined
wavelength. The low-temperature conductivity due to this band
reproduces some of the experimental phenomena in glasses, such as a linear
dependence on the temperature, and an eventual saturation (plateau).
In recent papers \cite{Allen}, it has been argued that the heat
transport in silicon glass is governed by harmonic modes;  our model
reproduces quite convincingly several of their results. In particular,
there is no  need to invoke non-linear processes such as two level
systems to explain the temperature dependence of the heat conductivity.
In all cases, however, the harmonic approximation yields a
conductivity which diverges as some small power of the system size,
and this feature must be corrected by the nonlinearity of the
interaction.

The disordered lattice model is thus found to be a useful tool to study
heat transport of glasses, and it will be very interesting to find out
what will 
happen with a vector vibration field and with random masses.  


{\bf Acknowledgments}: We have profited from helpful discussions with
I. Blanter, M. B\"uttiker, J.-P. Eckmann, G. Fagas, B. Galanti,
A. Politi, and R. Zeitak. Z. O. thanks Jean-Pierre Eckmann for the
invitation to the D\'epartement de Physique Th\'eorique of the University
of Geneva, which has been very useful for this work. This work has
been supported in part by the Fonds National Suisse.

\end{document}